\documentclass[twocolumn, noshowpacs,showkeys, amsmath,amssymb]{revtex4}

\newcommand{\ket}[1]{|#1\rangle}

\usepackage{graphicx}
\usepackage{dcolumn}
\usepackage{bm}
\usepackage{url}
\usepackage{natbib}
\usepackage{amsmath}
\usepackage{amssymb}
\begin{document}
\begin{abstract}
The stability of dynamical systems against perturbations (variations in initial conditions/model parameters) is a property referred to as \emph{structural stability}. The study of sensitivity to perturbation is essential because in experiment initial conditions are not fixed, nor are model parameters known, to arbitrarily high precision. Additionally, if a physical system under study exhibits stability (insensitivity to initial conditions) then a theoretical description of the system must exhibit structural stability. Consequently, stability can be a useful indicator of the correctness of a theoretical formulation. In this work the many-worlds interpretation is considered. It is first demonstrated that the interpretation admits a class of special states, herein referred to as ``quantum liar states," because they indicate disagreement between the recorded result of a measurement and the actual state of a system. It is then demonstrated that the many-worlds interpretation is not structurally stable against the introduction of quantum liar states.
 
\end{abstract}

\title{Structural Instability and Quantum Lying in the Many-worlds (Relative State) Interpretation}         
\author{Elliott Tammaro} 
\email{tammaroe@chc.edu}
\affiliation{Department of Physics, Chestnut Hill College, Philadelphia, Pennsylvania, USA}   
 \date{Received: date / Accepted: date}
\pacs{.23}
\maketitle

\section{Introduction}
\emph{Structural stability} is a property of dynamical systems which measures the sensitivity of the system to perturbation. It may be quantitatively defined in a variety of inequivalent ways depending on the type of perturbation and the measure of sensitivity. For example, one may consider altering model parameters or initial conditions and rating adjustments in short time or long time behavior. Or one may be concerned with quantitative versus qualitative differences upon perturbation. It must be emphasized that structural stability is a desirable component of an effective model. Low structural stability implies that seemingly minor changes vastly affect predictions. This, in turn, makes predictions less robust. Often physical systems exhibit stability. That is, their behavior is insensitive to small perturbations (in initial conditions for example). When this is the case any good theoretical description of the system must exhibit structural stability. In this manner stability can indicate the correctness (or at least viability) of a theoretical description. In this current work we concern ourselves with the many-worlds interpretation of quantum mechanics (MWI) as a possible resolution to the measurement problem. We do not offer a review of MWI here, as many good introductions can be found elsewhere in the literature, including the original relative state formulation due to Everett \cite{HughEverett1957}\cite{Bell}\cite{Tegmark}. We find it useful, as a matter of precision for both this work and future works, to define the term \emph{interpretation} as it is used in a physics context. We call a theory \emph{interpretable} and say that it has an \emph{interpretation} if statements made by the theory may be made to correspond to physically realizable situations. Consider for example classical mechanics wherein predictions of the theory correspond to functions $\mathbf{x}(t)$. An interpretation of the theory would take the functions $\mathbf{x}(t)$ to represent the trajectory of a particle. We see that classical mechanics has a clear interpretation and we refer to it as \emph{interpretationally trivial}. 
\section{Many-worlds interpretation: Quantum Liars/Stability}
Consider MWI. Na\"ive expectation dictates that as a viable interpretation of quantum mechanics MWI yields only statements that are interpretable (according to the aforementioned definition). That is, every statement made by the theory corresponds to some physically realizable situation. This is not the case however. To introduce just such a non-interpretable statement consider the quantum mechanical description of an observer + apparatus system $A$ and system of interest $S$. Any realistic observer + apparatus system would have a great many possible states. For example, the apparatus may read value $o$ while the observer is contemplating a trip to the grocery store. Or the apparatus may be ready to take a reading while the observer is sleeping. For simplicity we ignore such complications. We consider a system of interest whose basis states are discrete \begin{equation}\ket{o_1}_S,\ket{o_2}_S,\ldots\ket{o_n}_S.\end{equation} We assume that the apparatus outputs are $o_1,$ or $ o_2,$ or $o_3,\ldots$ or $o_n$ and that the observer merely makes a memory record of the output. As a basis for such an observer/apparatus system one may select states \begin{equation}\bigg\{ \ket{\mathrm{ready}}, \ket{o_1}_A, \ket{o_2}_A,\ldots\bigg\},\end{equation} where, naturally, $\ket{\mathrm{ready}}$ corresponds to the state in which the apparatus is prepared to make a measurement and the observer is waiting and states $\ket{o_i}_A$ correspond to the apparatus outputting $o_i$ and the observer making a memory record of said output. Consider now the following states of the full system (system of interest, observer, and apparatus) \begin{equation}\ket{o_i}_S\otimes\ket{o_j}_A \phantom{IIIII} i\neq j.\end{equation} 
Such states correspond to the observer/apparatus recording output $o_j$, while the system is actually in state $\ket{o_i}$. They, therefore, correspond to ``quantum lying" (or herein \emph{quantum liar states}) because the apparatus failed to the tell the truth about the state of the system of interest. Under the assumption that such states do not actually arise in the course of experiment we deem them ``non-interpretable." 

It is clear that quantum liar states arise in the state spectrum, but is it possible to merely banish them? To answer this question recall that the unitary representation of measurement in MWI is (essentially) that of a Von Neumann measurement of the first kind. That is to say, let a system of interest be initially in state \begin{equation}\ket{\psi}=\sum_i g_i\ket{o_i}.\end{equation} Then a measurement of the system is made by inducing an interaction $U_{M}$ which couples system of interest states with observer/apparatus states, i.e. \begin{equation}U_{M}\ket{\psi}\otimes \ket{\mathrm{ready}}=\sum_ig_i\ket{o_i}_S\otimes\ket{o_i}_A.\end{equation} 
Finally it is assumed that the environment interacts with system via unitary action $U_E$, which introduces decohering effects that prevent additional interference terms. That is, \begin{equation}U_{E}\left[\sum_ig_i\ket{o_i}_S\ket{o_i}_A\right]\otimes \ket{E_{in}}=\sum_ig_i\ket{o_i}_S\otimes\ket{o_i}_A\otimes\ket{E_i}.\end{equation} 
 Where we have introduced environment states $\ket{E_{in}}, \ket{E_1}, \ket{E_2},\ldots\ket{E_n}$\cite{HughEverett1957}\cite{ZurekI}\cite{ZurekII}\cite{ZurekIII}\cite{Stapp}\cite{Schlosshauer1}.

 It is evident that such a measurement scheme does not introduce quantum liar states. However, it is not structurally stable against their introduction. Consider, for example, the assumption that the observer/apparatus system is found initially in $\ket{\mathrm{ready}}$. It is apparent that more general initial states exist. Indeed, generically the initial state could be \begin{equation}\alpha\ket{\mathrm{ready}}+\beta_1\ket{o_1}+\beta_2\ket{o_2}+\ldots\beta_n\ket{o_n} \label{b},\end{equation} with arbitrary normalized coefficients $\alpha, \beta_1,\beta_2,\ldots\beta_n$. The effect of a more generic initial observer/apparatus state such as \eqref{b} is to introduce quantum liar states. Let us demonstrate this by considering the following initial state \begin{equation}\alpha\ket{\mathrm{ready}}+\beta\ket{o_k}.\end{equation}
Although not the most general initial state it suffices to make our point.
The time evolution is as follows \begin{eqnarray}U_{M}\ket{\psi}\otimes( \alpha\ket{\mathrm{ready}}&+&\beta\ket{o_k})= \alpha\sum_ig_i\ket{o_i}_S\otimes\ket{o_i}_A \nonumber\\&+&\beta\sum_ig_iU_M(\ket{o_i}_S\otimes\ket{o_k}_A).\label{1}\end{eqnarray} The first term in equation \eqref{1}, $\alpha\sum_ig_i\ket{o_i}_S\otimes\ket{o_i}_A$, is expected (interpretable) as it represents perfect coupling between observer/apparatus and system of interest. Observe that the second term, $\beta\sum_ig_iU_M(\ket{o_i}_S\otimes\ket{o_k}_A)$, depends on the unitary action of $U_M$ on states $\ket{o_i}_S\otimes\ket{o_k}_A$, where $k$ is arbitrary. $U_M(\ket{o_i}_S\otimes\ket{o_k}_A)$ is not known a priori, but depends instead on the particular choice for $U_M$. There are three cases to consider. The most desirable possibility, that $U_M(\ket{o_i}_S\otimes\ket{o_k}_A)$ vanishes, cannot occur because such a nontrivial vanishing implies a loss of unitarity. A second possibility is that $U_M(\ket{o_i}_S\otimes\ket{o_k}_A)$ contains only perfect coupling terms such as $\ket{o_i}_S\otimes\ket{o_i}_A$. If this were so then the second term would make contributions to the first thereby altering the coefficients $g_i$. In turn this would affect probabilities computed via the Born rule, which is empirically well verified, or equivalently, it would alter the final state after decohering effects occur. Additionally, it would be hard to construct a unitary operator $U_M$ to accomplish this feat. Finally, it is possible that $U_M\left(\ket{o_i}_S\otimes\ket{o_i}\right)$ yields terms of the form $\ket{o_i}_S\otimes\ket{o_j}$ with $i\neq j$, which are the aforementioned quantum liar states, as was to be shown.

Are quantum liar states generated in experiment? While possible, we believe it is unlikely that quantum liar states are generated. It is noteworthy that making repeat measurements on a single system could reveal them. Namely, if a system is determined to be in eigenstate $\ket{o_i}$ then a repeat measurement will confirm it (i.e. a repeat measurement will continue to yield eigenvalue $o_i$). If quantum liar states are generated then it is possible, let us say, for the first measurement to reveal $o_a$ when the system was actually in $\ket{o_b}$, which upon a second (good) measurement would be revealed as $\ket{o_b}$. Thusly, as long as the eigenstate-eigenvalue link continues to be confirmed the possibility of quantum liar states is exceedingly small. This raises an obvious question, however, if quantum liar states are not generated, then where are they? That is, no known mechanism \emph{prevents} the formation of initial states that would give rise to quantum liars. Indeed, it is unrealistic to assume that an observer/apparatus system, which must be a very rich quantum system, could \emph{always} be prepared in precisely the correct initial state in order to eliminate quantum liars. Hence, it is necessary to conclude that MWI, without the introduction of additional mechanisms, contains an instability which leads to the introduction of quantum liars.

 We note in passing that decoherence approaches which make use of Von Neumann measurements of the first kind also suffer from an instability against the introduction of quantum liar states. More precisely, the environment is often modeled as system with states \begin{equation}\bigg\{\ket{E_{in}}, \ket{E_1},\ket{E_2},\ldots \ket{E_n}\bigg\}\end{equation} where $\ket{E_{in}}$ is the initial environment state and $\ket{E_i}$ are possible ``out" states, which correspond to particular results of measurement. In such decoherence models the final state is assumed to be the pure coupling state \begin{equation}\sum_ig_i\ket{o_i}_S\otimes\ket{o_i}_A\otimes\ket{E_i}.\end{equation} Thusly, interaction with the environment could produce quantum liar states \begin{eqnarray}\ket{o_i}_S\otimes\ket{o_i}_A\otimes\ket{E_j}\phantom{IIIIII}i\neq j \end{eqnarray} unless the environment is prepared precisely in $\ket{E_{in}}$. There does not appear to be an obvious mechanism that would guarantee that the initial environment state is $\ket{E_{in}}$. A similar observation was noted in \cite{Tammaro} wherein it is referred to as the \emph{initial entropy problem} \cite{ZurekI}\cite{ZurekII}\cite{ZurekIII}\cite{Schlosshauer1}\cite{Schlosshauer2}\cite{Zeh}.

\section{Ackowledgments}
The author would like to acknowledge T. Chard for several illuminating conversations.

\bibliographystyle{unsrtnat}

\end{document}